\documentclass[epj]{svjour}
\usepackage{graphics}
\usepackage{bm}   
\usepackage{color}
\usepackage{amssymb}
\usepackage{amsmath}
\usepackage{balance}

\newcommand{\e}{\normalfont\mbox{e}\,}

\begin{document}

\title{On the characterization of vector rogue waves in two-dimensional two coupled nonlinear Schr\"{o}dinger equations with distributed coefficients}
\author{K. Manikandan \inst{1} M. Senthilvelan \inst{1,} \thanks{\emph{e-mail:velan@cnld.bdu.ac.in}} \and R.A. Kraenkel \inst{2}}

\institute{Centre for Nonlinear Dynamics, Bharathidasan University, Tiruchirappalli 620024, Tamilnadu, India \and Instituto de F\'{\i}sica Te\'orica , Universidade Estadual Paulista, Rua Dr. Bento Teobaldo Ferraz 271, 01140-070 S\~ao Paulo, Brazil}

\abstract{
We construct vector rogue wave solutions of the two-dimensional two coupled nonlinear Schr\"{o}dinger equations with distributed coefficients, namely diffraction, nonlinearity and gain parameters through similarity transformation technique. We transform the two-dimensional two coupled variable coefficients nonlinear Schr\"{o}dinger equations into Manakov equation with a constraint that connects diffraction and gain parameters with nonlinearity parameter.  We investigate the characteristics of the constructed vector rogue wave solutions with four different forms of diffraction parameters. We report some interesting patterns that occur in the rogue wave structures. Further, we construct vector dark rogue wave solutions of the two-dimensional two coupled nonlinear Schr\"{o}dinger equations with distributed coefficients and report some novel characteristics that we observe in the vector dark rogue wave solutions. 
}
\titlerunning{Vector rogue waves in two-dimensional two coupled vcNLS equations}
\authorrunning{K. Manikandan et al.}
\maketitle

\section{Introduction}
A rogue wave (RW) is a nonlinear wave which is localized in both space and time and appears from nowhere and disappears without a trace \cite{osbrn:rato}.  It was first observed in ocean \cite{osbrn} with a wave-height (the distance from trough to crest) two or three times greater than the significant wave height \cite{osbrn,khar:pelin}.  It arises due to the instability of a certain class of initial conditions that tend to grow exponentially and thus have the possibility of increasing up to very high amplitudes, due to modulation instability.  Later RWs have been observed in many physical systems including water wave tank experiments \cite{chab}, capillary waves \cite{shatz}, nonlinear optics \cite{solli,kibler} and Bose-Einstein condensates \cite{kono}.  Subsequently, efforts have been made to explain the RW excitation through nonlinear process.  It has been found that the nonlinear Schr\"{o}dinger (NLS) equation can describe the structural and dynamical properties of the RWs. A rational solution of the NLS equation can be used to model the RWs \cite{pere,akmv:anki}.  During the past ten years or so several works have been devoted to study and analyze the characteristics of RW solutions in the NLS equation and its variants by considering their potential applications in transmitting highly intense signals through optical fibers. For a summary of recent progress on optical RWs we refer to \cite{akhm} and the references therein.  

In real fibers, there always exist some nonuniformities owing to diverse factors such as imperfections of manufacture, variations in the lattice parameters of the fiber media, fluctuations in the fiber diameters and so on \cite{book1}. These nonuniformities often produce fiber gain or loss, phase modulation and variable dispersion \cite{book1}.  The variable coefficients NLS (vcNLS) equation is an efficient model to study the inhomogeneous effects of nonlinear optical pulses \cite{book2}. Unlike the constant coefficient NLS equation the studies on vcNLS equation reveal that one can control/amplify the localized structures through their inhomogeneity parameters.  Subsequently, identifying and the possibility of controlling RWs in the one-dimensional vcNLS equations have been investigated by several authors \cite{serk,yan,wen:li,kumar,Mani}.  Even though the one-dimensional equations have given a good understanding on the dynamics,  a detailed investigation on the higher-dimensional version (two or three dimensions) of the system will provide a clear visualization about the localized structures.  Motived by this, few investigations have been made on the higher dimensional versions of scalar vcNLS equations.  For example, (i) self-similar solutions of the higher dimensional vcNLS equation have been obtained analytically and numerically in \cite{dswang,zyan,zjfang,zyan2,mani3,mani4}, (ii) stable bright and vortex solitons for a two-dimensional NLS equation have been identified through numerical simulations in \cite{adhi}, (iii) exact bright and dark similariton pairs have been identified in a generalized two-dimensional NLS equation with distributed coefficients in \cite{cqdai} and (iv) nonlinear tunneling of controllable RWs in the two-dimensional graded-index waveguides have been analyzed in \cite{zhu,dai}. 

In certain physical situations, two or more wave packets of different frequencies or polarizations appear simultaneously and their interactions are then governed by the coupled NLS equations \cite{book1,book2,book3}. For example, nonlinear light propagation in a birefringent optical fiber or a wavelength-division-multiplexed system \cite{book2,book3}, the evolution of two surface wave packets in deep water \cite{rosk}, the interaction of Bloch-wave packets in a periodic system \cite{shi} and spinor Bose-Einstein condensates \cite{dalf}. All the above processes are modelled by coupled NLS equations.  These vector systems allow energy transfer between their additional degress of freedom.  The vector NLS equations yield potentially rich and significant results for optical fiber communication systems \cite{book1}.  Motivated by this, several studies have also been undertaken to identify the localized solutions in coupled NLS equations with constant coefficients \cite{mana,radha,radha1,kanna,fabio,vishnu,ling,lomb,zhai,lomb1,vishnu3}.  Subsequently attempts have been made to identify the localized solutions such as bright-bright, bright-dark, dark-dark soliton and RW solutions in the coupled vcNLS equations (in one-dimensions) \cite{raje,rama,vina,babu,fyu,Mani2,cheng,wpzhong}. Very recently attempts have been made to identify RW and breather solutions in the (2+1)-dimensional coupled vcNLS equations \cite{dai2,kkde}.  To construct these localized structures the authors of \cite{dai2,kkde} have transformed the (2+1)-dimensional coupled vcNLS equations into scalar (1+1)-dimensional NLS equation. As we pointed out earlier the coupled constant coefficient NLS equations exhibit certain new RW structures that are absent in the scalar NLS equation.  In other words transforming a (2+1)-dimensional coupled vcNLS equations into the coupled constant coefficient NLS equations will reveal certain new localized solutions, vector RWs and dark-dark RWs, for the equation under investigation. Motivated by this, in this paper, we transform the two coupled vcNLS equations into two coupled constant coefficient NLS equations.  Through this approach we construct vector RW solutions and study their characteristics in the two-dimensional two coupled NLS equations with distributed coefficients such as diffraction, nonlinearity and gain/loss parameter. 

In this work, we transform the considered equation into two component NLS equation (Manakov equation) through the similarity transformation with a constraint that connects diffraction and gain parameters with nonlinearity parameter.  From the known RW solutions of the Manakov equation we derive RW solutions of the two-dimensional two coupled vcNLS equations. The Manakov equation exhibits two different forms of RW solutions \cite{ling}.  These two forms in turn yield two different types of RW solutions for the two-dimensional two coupled vcNLS equations.  We investigate the characteristics of these two vector RWs with four different forms of diffraction parameters, namely (i) $\beta(z)=\beta_0 \cos{(\sigma z)}\exp{(\gamma_0 z)}$, (ii) $\beta(z)=\beta_1 \exp{(-\beta_0 z)}$, (iii)  $\beta(z)=\beta_1-\beta_0 z$ and (iv) $\beta(z)=\beta_1 -\beta_0 z-\beta_2 z^2/2$, where $\beta_0$, $\beta_1$ and $\beta_2$ are positive constants.  We consider the same gain profile, namely $\gamma(z)=\gamma_0/2$ in all the four cases. Once the diffraction and gain parameters are chosen the nonlinearity parameter can be fixed through the constraint.  For a constant diffraction parameter, the vector RWs show usual RW features but are localized in different orientations in each component.  In the case of exponentially growing periodic diffraction parameter (case (i)) the RWs propagate periodically and their amplitudes increase along the propagation direction in both the components. In the exponentially distributed diffraction parameter case (case (ii)) the amplitude of the RWs monotonically increases along the propagation direction in both the components.  In case (iii) we come across a composite RWs (RW pair) in each component.  The RWs merge together when we increase the value of the parameter $\beta_0$ and they separate out in the plane when we decrease the value of this parameter.  In the fourth profile we observe three separate RWs (composite RWs) that occur at three different positions along the propagation direction.  

Next we move on to investigate the characteristics of the dark-dark RWs in the (2+1)-dimensional coupled vcNLS equations.  In this equation the nonlinearity and the diffraction parameters are in opposite signs which in turn produce a new localized structure, namely dark RW.  Recently, several works on  dark RWs have been made by considering their potential applications in fiber optic communications \cite{lomb,lomb1,jhli}.  The dark RW differs from the Peregrine soliton.  In the dark RW the amplitude at the centre has less value than the amplitude of the background \cite{jhli}. We construct vector dark RW solutions for the (2+1)-dimensional two coupled vcNLS equations by transforming them into the two coupled constant coefficient NLS equations. The transformed equation is different from the one which we considered to derive the bright RWs above.  From the known dark RW solutions of the latter equation we construct dark RW solutions of the considered two coupled vcNLS equations.  We then examine the characteristics of dark-dark RWs with the same diffraction parameters that we considered in the bright RW case. In the constant diffraction parameter case, the dark RW in each component reveal the usual RW features whereas in the exponentially growing periodic diffraction parameter case (case (i)) the dark RWs exhibit breathing profiles along the propagation direction in both the components.  As far as the exponentially distributed diffraction parameter is concerned (case (ii)) the dip of the modified dark RWs become more darkened in both the components.  In the third profile, an interaction between two dark RWs (composite RWs) occurs in both the components in the $(x-z)$ plane. In the fourth profile, an interaction between three dark RWs (composite RWs) occur along the propagation direction in both the components. When we increase the value of the gain parameter the background gets more steepened and if we decrease the value of the gain parameter, the background becomes flattered in both the components. 

We organize our work as follows.  In Section 2, we consider the (2+1)-dimensional two coupled vcNLS equations and construct the vector RW solutions of it.  In Section 3, we investigate in detail the characteristics of vector RWs by considering four different forms of diffraction parameters.  In Section 4, we construct the vector dark RW solutions of (2+1)-dimensional two coupled vcNLS equation and investigate the characteristics of them by considering the same forms of diffraction parameters. In Section 5, we present our conclusions. 
\section{Model and RW solutions of two-dimensional two coupled vcNLS equations}
Multimode wave propagation in inhomogeneous nonlinear waveguides are described by the following two-dimensional two coupled vcNLS equations \cite{book1,book2}, that is 
\begin{align}
i\psi_{1z}+\frac{\beta(z)}{2}\left(\psi_{1xx}+\psi_{1yy}\right)+R(z)\sum_{k=1}^{2} \vert \psi_k \vert^2 \psi_1 = i \gamma(z)\psi_1, \nonumber \\
i\psi_{2z}+\frac{\beta(z)}{2}\left(\psi_{2xx}+\psi_{2yy}\right)+R(z)\sum_{k=1}^{2} \vert \psi_k \vert^2 \psi_2 = i \gamma(z)\psi_2,
\label{cdis:eq1}
\end{align}
where $\psi_j(x,y,z)$, $j=1,2$, represent complex envelope of the electrical field, $z$ represents the normalized propagation distance along the waveguides and $x$ and $y$ are the transverse coordinates.  The functions $\beta(z)$, $R(z)$ and $\gamma(z)$ denote the diffraction, nonlinearity and gain/loss parameter, respectively and all of them are real analytic functions of $z$. To identify the vector RW solutions and investigate the dynamical evolutions of them in (\ref{cdis:eq1}), we consider a similarity transformation of the form 
\begin{eqnarray}
\psi_j(x,y,z)=\rho(z)U_j(X,Z)\exp[i \phi(x,y,z)], \;\; j=1,2,
\label{cdis:eq2}
\end{eqnarray}
where $\rho(z)$ is the amplitude, $Z(z)$ is the effective propagation distance, $X(x,y,z)$ is the similarity variable and $\phi(x,y,z)$ is the phase factor which are all to be determined. We also demand that the ~complex ~functions ~$U_j(X,Z)$ should satisfy ~the two~ component NLS equations of the form 
\begin{eqnarray}
\label{cnls}
i \frac{\partial U_1}{\partial Z}+\frac{\partial ^2 U_1}{\partial X^2}+ 2 (|U_1|^2+|U_2|^2) U_1=0, \nonumber \\
i \frac{\partial U_2}{\partial Z}+\frac{\partial ^2 U_2}{\partial X^2}+ 2 (|U_1|^2+|U_2|^2) U_2=0,
\end{eqnarray}
whose solutions are known.  To determine the unknown functions of (\ref{cdis:eq2}) we substitute it into (\ref{cdis:eq1}) and obtain the following set of partial differential equations (PDEs) for these unknown functions, that is
\begin{eqnarray}
&& X_{xx}+X_{yy}=0, \notag \\
&& X_z+\beta(z)(X_x\phi_x+X_y\phi_y)=0, \notag \\ 
&& \phi_z+\frac{\beta(z)}{2}(\phi_x^2+\phi_y^2)=0, \label{pdes} \\
&& \frac{\rho_z}{\rho(z)}+\frac{\beta(z)}{2}\left(\phi_{xx}+\phi_{yy}\right)-\gamma(z)=0,  \notag\\
&& Z_z-R(z)\rho^2(z)=0, \;\;\; \beta(z)(X_x^2+X_y^2)-R(z)\rho^2(z)=0. \notag 
\end{eqnarray}
Solving these PDEs, we find the following expressions, namely  
\begin{subequations}
\begin{align}
\label{cdis:eq3a}  q(z)=& \, \frac{q_0}{f(z)}, \;\; r(z)= \frac{r_0}{f(z)}, \\
\label{cdis:eq3a1} s(z)=& \, \frac{-s_0+(b_0+a_0 s_0) M(z)}{f(z)}, \\
\label{cdis:eq3b} X(x,y,z)= & \, q(z) x + r(z) y+ s(z),  \\
\label{cdis:eq3b1} G(z) = & \, \int_0^z\gamma(s)ds, \\
\label{cdis:eq3c} Z(z)= & \, \frac{(q_0^2+r_0^2)M(z)}{2f_0^2(1-a_0 M(z))}, \\
\label{cdis:eq3c1} \rho(z)= & \, \frac{\rho_0 \exp{(G(z))}}{1-a_0 M(z)}, \\
\label{cdis:eq3d} \phi(x,y,z)=& \, \phi_0-\frac{a_0(x^2+y^2)}{2(1-a_0 M(z))}-\frac{b_0(x/q_0+y/r_0)}{2(1-a_0 M(z))} \notag \\
                              & \, -\frac{(q_0^2+r_0^2)b_0^2 M(z)}{8q_0^2r_0^2(1-a_0 M(z))},
\end{align}
\end{subequations}
where $M(z)=\int_0^z\beta(s)ds$ represents the accumulated dispersion/diffraction and the function $f(z)=f_0(1-a_0M(z))$ can be related to the wavefront curvature of the wave \cite{dai}. The parameters $a_0$, $b_0$, $\rho_0$, $s_0$, $f_0$, $q_0$ and $r_0$, which arise from integration, can be interpreted as follows: The parameters $a_0$ and $b_0$ are the initial curvature and position of the wavefront, $\rho_0$ and $s_0$ are the initial amplitude and position of the pulse center, $f_0$ is the initial wave profile width and $q_0$ and $r_0$ are the group velocity parameters, respectively \cite{dai}.  

The existence of localized structures in (\ref{cdis:eq1}) can be guaranteed by the fulfillment of the following constraint between gain/loss and diffraction parameter with the nonlinearity parameter, that is
\begin{equation}
\label{cdis:eq4}
R(z)=\frac{\beta(z)(q_0^2+r_0^2)\exp{(-2G(z))}}{(f_0\rho_0)^2},
\end{equation}
where $G(z)$ is given in (\ref{cdis:eq3b1}).  We can choose $\gamma(z)$ and $\beta(z)$ arbitrarily but the nonlinearity parameter $R(z)$ should be fixed as per (\ref{cdis:eq4}).  Once the constraint (\ref{cdis:eq4}) has been accommodated the solutions of (\ref{cdis:eq1}) can be picked up from the solutions of the two component NLS equation (\ref{cnls}). 

Equation (\ref{cnls}) is the well-known Manakov equation whose integrability properties have been extensively studied in the literature \cite{mana,radha,radha1,kanna,fabio,vishnu,ling,lomb,zhai,lomb1}. A detailed investigation on the soliton solutions of two coupled NLS equation with same nonlinearity (nonlinear coefficients have the same signs '+' and '-') and mixed nonlinearity (nonlinear coefficients have opposite signs) gave birth to three different vector soliton solutions, namely (i) bright-bright \cite{mana,radha}, (ii) bright-dark \cite{shep,viji} and (iii) dark-dark solitons \cite{radha,prin}.  The explicit multi-bright and multi-dark soliton solutions of the Manakov system were obtained in \cite{radha,radha1}.  Recently, one dark-dark soliton and the general breather solution of a generalized version of Eq. (\ref{cnls}) have been constructed through Hirota's bilinearization method \cite{vishnu2}.  Very recently, a new form of localized solution, called RW solution, has attracted considerable attention in the two coupled NLS equation (\ref{cnls}) \cite{fabio,vishnu,lomb,ling}. In \cite{lomb,zhai,lomb1}, explicit RW solutions of two coupled NLS equation have been constructed through modified Darboux transformation method.  Interestingly two types of RW solutions have been obtained for (\ref{cnls}) through Darboux transformation method \cite{ling}.  Since we are interested in constructing vector RW solutions of (\ref{cdis:eq1}) we consider both of them in our analysis \cite{ling}.  

Equation (\ref{cnls}) admits two different forms of RW solutions which we call Type-I RW and Type-II RW, respectively.  The exact form of Type-I RW is given by \cite{ling} 
\begin{eqnarray}
\label{soln}
U_1(X,Z)=& \alpha\exp(i\theta_1)\left(-1-i\sqrt{3} \right. \qquad \qquad \qquad \qquad \notag \\ & \left. +\frac{-6\delta\alpha\sqrt{3}-36Z\alpha^2\sqrt{3}-3+i(36\alpha^2 Z+6\delta \alpha+5\sqrt{3})}{12\alpha^2\delta^2+8\delta\alpha\sqrt{3}+144Z^2\alpha^4+5}\right),  \nonumber \\
U_2(X,Z)=& \alpha\exp(i\theta_2)\left(-1+i\sqrt{3}\right. \qquad \qquad \qquad \qquad \notag \\  & \left. +\frac{-6\delta\alpha\sqrt{3}+36Z\alpha^2\sqrt{3}-3+i(36\alpha^2 Z-6\delta \alpha-5\sqrt{3})}{12\alpha^2\delta^2+8\delta\alpha\sqrt{3}+144Z^2\alpha^4+5}\right), 
\end{eqnarray}
where $\delta =X+6 p Z$, $\theta_1=d_1 X+(2 c_1^2+2c_2^2-d_1^2) Z$ and $\theta_2=d_2 X+(2 c_1^2+2c_2^2-d_2^2) Z$, respectively. The parameters $\alpha=d_2+3p$, $d_1=d_2-2\alpha$, $c_1, c_2=\pm 2\alpha$, $d_2$ and $p$ are arbitrary constants. The RW solution given above is similar to the first-order RW solution of the scalar NLS equation \cite{akmv:anki}. 

Type-II RW solution is given by 
\begin{subequations}
\label{combsoln}
\begin{eqnarray}
\label{csoln}
U_1(X,Z)=\alpha\left(-1-i\sqrt{3}+\frac{G_1+iH_1}{D}\right)\exp(i\theta_1), \nonumber \\
U_2(X,Z)=\alpha\left(-1+i\sqrt{3}+\frac{G_2+iH_2}{D}\right)\exp(i\theta_2),  
\end{eqnarray}
where
\begin{eqnarray}
D & = & 1+4\sqrt{3}\alpha\delta+24\alpha^2\delta^2+16\sqrt{3}\alpha^3\delta^3+12\alpha^4\delta^4 \nonumber \\ && +48\alpha^4(9+8\sqrt{3}\alpha\delta+6\alpha^2\delta^2)Z^2+1728\alpha^8Z^4, \nonumber \\
G_1 &= & -3(-1+6\alpha^2\delta^2+4\sqrt{3}\alpha^3\delta^3+4\alpha^2(\sqrt{3}+12\alpha\delta \nonumber \\ && +6\sqrt{3}\alpha^2\delta^2)Z +24\alpha^4(3+2\sqrt{3}\alpha\delta)Z^2 +288\sqrt{3}\alpha^6Z^3), \nonumber \\
G_2&= & 3(1-6\alpha^2\delta^2-4\sqrt{3}\alpha^3\delta^3+4\alpha^2(\sqrt{3}+12\alpha\delta \nonumber \\ && +6\sqrt{3}\alpha^2\delta^2)Z -24\alpha^4(3+2\sqrt{3}\alpha\delta)Z^2  + 288\sqrt{3}\alpha^6Z^3), \nonumber \\
H_1&= & \sqrt{3}+12\alpha\delta+18\sqrt{3}\alpha^2\delta^2+12\alpha^3\delta^3+12\alpha^2(9+8\sqrt{3}\alpha\delta\nonumber \\ &&+6\alpha^2\delta^2)Z +24\alpha^4(13\sqrt{3} +6\alpha\delta)Z^2+864\alpha^6Z^3, \\
H_2&= &-\sqrt{3}-12\alpha\delta-18\sqrt{3}\alpha^2\delta^2-12\alpha^3\delta^3 \nonumber \\ &&+12\alpha^2(9  +8\sqrt{3}\alpha\delta+6\alpha^2\delta^2)Z -24\alpha^4(13\sqrt{3}  \nonumber \\ && +6\alpha\delta)Z^2+864\alpha^6Z^3 \nonumber.
\end{eqnarray}
\end{subequations}
The strucutral difference between Type-I and Type-II RW solutions is as follows. Type-I RW  solution is similar to the Peregrine soliton which has one largest crest and two troughs as shown in Fig. \ref{cdis:fig1}(a) whereas Type-II RW solution has one largest crest, two subcrests and two troughs, as shown in Fig. \ref{cdis:fig2}(a).  

By choosing a suitable form of diffraction parameter $\beta(z)$ and the gain/loss profile $\gamma(z)$ and considering the nonlinearity parameter which comes out from the condition (\ref{cdis:eq4}), the exact solution of Eq. (\ref{cdis:eq1}) can now be produced in the form  
\begin{eqnarray}
\label{cdis:eq6}
\psi_j(x,y,z)&=&\frac{\rho_0\exp(G(z))}{1-a_0M(z)} U_j(X,Z)  \exp \left[\left\{i\left(\phi_0 \right. \right. \right. \nonumber \\ && \left. \left. \left. -\frac{a_0(x^2+y^2)}{2(1-a_0M(z))}  -\frac{b_0(x/q_0+y/r_0)}{2(1-a_0M(z))} \right. \right. \right. \nonumber \\ && \left. \left. \left. -\frac{(q_0^2+r_0^2)b_0^2M(z)}{8q_0^2r_0^2(1-a_0M(z))}\right)\right\} \right], \;\; j=1,2,
\end{eqnarray}
where $U_j(X,Z)'$s are given in Eqs. (\ref{soln}) and (\ref{combsoln}). We identify certain novel optical RW structures of (\ref{cdis:eq1}) from (\ref{cdis:eq6}) by selecting the arbitrary functions appropriately. The arbitrary functions will provide us further choices to develop fruitful structures related to the optical RWs, which may be useful to raise the feasibility of corresponding experiments and potential applications of optical pulses in real world communication systems.  
\section{Characteristics of vector RWs in two-dimensional two coupled vcNLS equations}
To understand the behaviour of the obtained exact solution (\ref{cdis:eq6}) of (1), we consider certain specific forms for diffraction and gain parameters. The nonlinearity parameter should be fixed through the constraint (\ref{cdis:eq4}). With these forms we analyze the solution profile (\ref{cdis:eq6}). We will consider four different forms of diffraction parameters and study the characteristics of the associated inhomogeneous RWs.
\subsection{Case 1}
Periodic distributed systems are found potential applications in long-distance communication systems \cite{maha}.  Motivated by this, to begin, we consider the diffraction parameter $\beta(z)$ be an exponentially growing periodic diffraction in $z$, that is $\beta(z)=\beta_0 \cos{(\sigma z)}\exp{(\gamma_0 z)}$ and the gain parameter as $\gamma(z)=\gamma_0/2$.  Here $\sigma$ is related to the variation period of the waveguide parameter and $\gamma_0$ is an arbitrary parameter \cite{cqdai}. With these parameters the solution reads now
\begin{subequations}
\label{cdis:eq7}
\begin{align}
\psi_j(x,y,z) = & \, \frac{\rho_0(\gamma_0^2+\sigma^2)}{\gamma_0^2+\sigma^2-\e^{\gamma_0 z}a_0\beta_0s_2}\notag \\ & \, \times U_j(X,Z)  \eta(x,y,z), \;\; j=1,2,
\end{align}
\begin{align}
\eta(x,y,z) = & \exp\left\{\frac{\gamma_0 z}{2}+\frac{\e^{\gamma_0 z}\beta_0s_2((q_0^2+r_0^2)b_0^2+8q_0^2r_0^2a_0\phi_0)}{8q_0^2r_0^2(\gamma_0^2+\sigma^2-\e^{\gamma_0 z}a_0\beta_0s_2)} \right. \notag \\ & \left. -\frac{s_1(q_0 b_0 y+r_0(b_0 x+q_0 a_0(x^2+y^2)-2q_0\phi_0))}{8q_0^2r_0^2(\gamma_0^2+\sigma^2-\e^{\gamma_0 z}a_0\beta_0s_2)}\right\},
\end{align}
\end{subequations}
where $s_1=4 q_0r_0(\gamma_0^2+\sigma^2)$, $s_2= \gamma_0\cos{(\sigma z)} +\sigma\sin{(\sigma z)}$, and $U_j(X,Z)$'s are given in Eqs. $(\ref{soln})$ and $(\ref{combsoln})$, respectively. 
\begin{figure*}[!ht]
\begin{center}
\resizebox{0.75\textwidth}{!}{\includegraphics{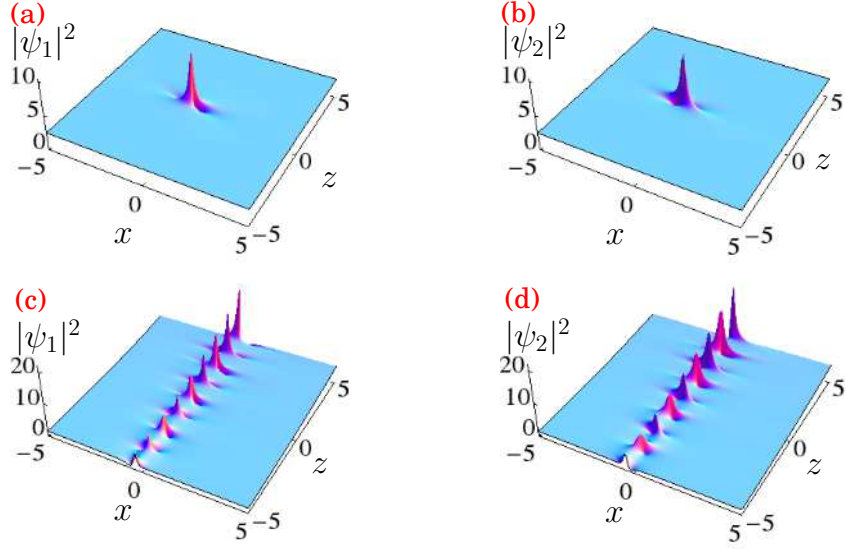}}
\end{center}
\caption{Intensity profiles of Type-I RWs for $\beta(z)=\beta_0 \cos{(\sigma z)}\exp{(\gamma_0 z)}$ and $\gamma(z)=\gamma_0/2$.  When the parameter $\gamma_0 =0$ and $\sigma =0$ for (a)-(b) and $\gamma_0=0.15$ and $\sigma=2.5$ for (c)-(d).  The other parameters are fixed as $q_0 ,r_0, \rho_0=1.0$, $s_0=0.01$, $\beta_0=f_0=0.5$, $a_0=b_0=0.1$, $y=0.1$, $p=\phi_0=0.01$ and $d_2=0.5$.}
\label{cdis:fig1}
\end{figure*}
\begin{figure*}[!ht]
\begin{center}
\resizebox{0.75\textwidth}{!}{\includegraphics{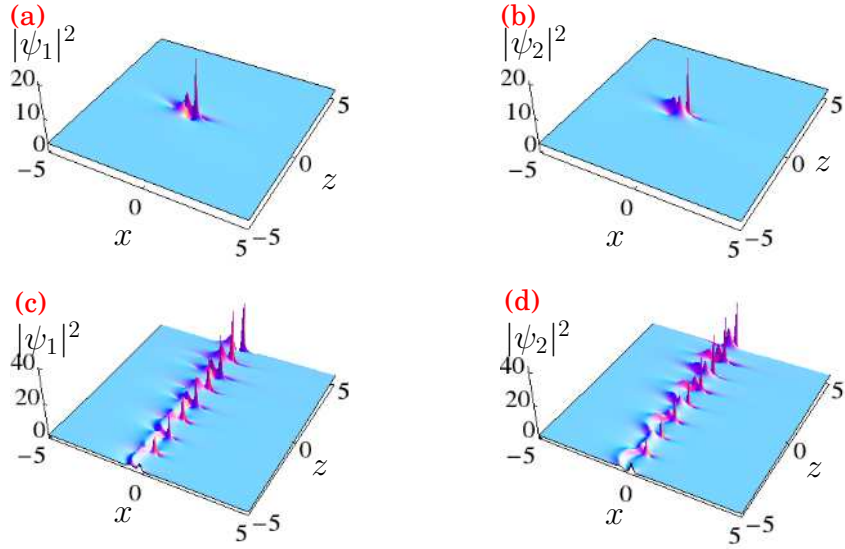}}
\end{center}
\caption{Intensity profiles of Type-II RWs for $\beta(z)=\beta_0 \cos{(\sigma z)}\exp{(\gamma_0 z)}$ and $\gamma(z)=\gamma_0/2$.  When the parameter $\gamma_0 =0$ and $\sigma =0$ for (a)-(b) and $\gamma_0=0.1$ and $\sigma=2.5$ for (c)-(d).  The other parameters are same as in Fig. \ref{cdis:fig1}}.
\label{cdis:fig2}
\end{figure*}

Fig.~\ref{cdis:fig1} shows the intensity profiles of Type-I RWs of (\ref{cdis:eq1}) obtained from (\ref{cdis:eq7}) with the diffraction parameter $\beta(z)=\beta_0 \cos{(\sigma z)}\exp{(\gamma_0 z)}$. We begin our study by considering the diffraction parameter is a constant, that is $\beta(z)=\beta_0$.  The corresponding intensity profiles of RWs are shown in Figs.~\ref{cdis:fig1}(a)-\ref{cdis:fig1}(b) where one can see the usual RW features.  The amplitudes of RWs are equal but they are localized in different positions in the $(x-z)$ plane.  We choose $\gamma_0=0.1$ and $\sigma=2.5$ and depict the intensity profiles in Figs.~\ref{cdis:fig1}(c)-\ref{cdis:fig1}(d).  We observe that the RWs breathe along the propagation direction and their amplitudes are enhancing in the propagation direction which can induce rapid periodic changes in the refractive index of the waveguide. The breathing comes out due to the presence of a periodic function in $\beta(z)$.  The total number of photons of the system also increase continuously along $z$ due to the presence of gain term (\ref{cdis:eq1}).  The outcome reveals that the system (\ref{cdis:eq1}) is not a conservative one.  This nonconservative nature of system causes growth in the amplitude of the RW.  The addition of photons (through the gain term) does not destabilize the localized structure of the system.  The width of the RWs also increases along the propagation direction due to the presence of exponential factor in $\beta(z)$.  

Substituting (\ref{combsoln}) into (\ref{cdis:eq7}), we obtain Type-II RW solution for (\ref{cdis:eq1}).  This RW solution replicates a beam propagating in nonlinear waveguides. The corresponding intensity profiles are shown in Fig.~\ref{cdis:fig2}.  The intensity profiles for the constant diffraction parameter $\beta(z)=\beta_0$ are shown in Figs.~\ref{cdis:fig2}(a)-\ref{cdis:fig2}(b).  The amplitudes of the RWs are equal but they are localized in different orientations (in particular the two subcrests of RWs) in the $(x-z)$ plane.  As we done in the previous case we choose the parameters as $\gamma_0=0.1$ and $\sigma=2.5$ and study the changes that occur in the RW pattern.  The intensity profiles exhibit breathing profiles and their amplitudes increase along the propagation direction as shown in Figs.~\ref{cdis:fig2}(c)-\ref{cdis:fig2}(d).  

From the above observations, we conclude that both the Type-I and Type-II RWs do not get distorted in the constant diffraction parameter whereas for exponentially growing periodic diffraction parameter both the Type-I and Type-II RWs exhibit breathing behaviour and their amplitudes grow in the propagation direction. Thus, RWs can be amplified along the propagation distance without any other relay device because of the inhomogeneous nature of optical waveguides. This amplification process do not introduce noise and the quality of optical amplification can be improved through this manner \cite{wjliu}.
\begin{figure*}[!ht]
\begin{center}
\resizebox{0.75\textwidth}{!}{\includegraphics{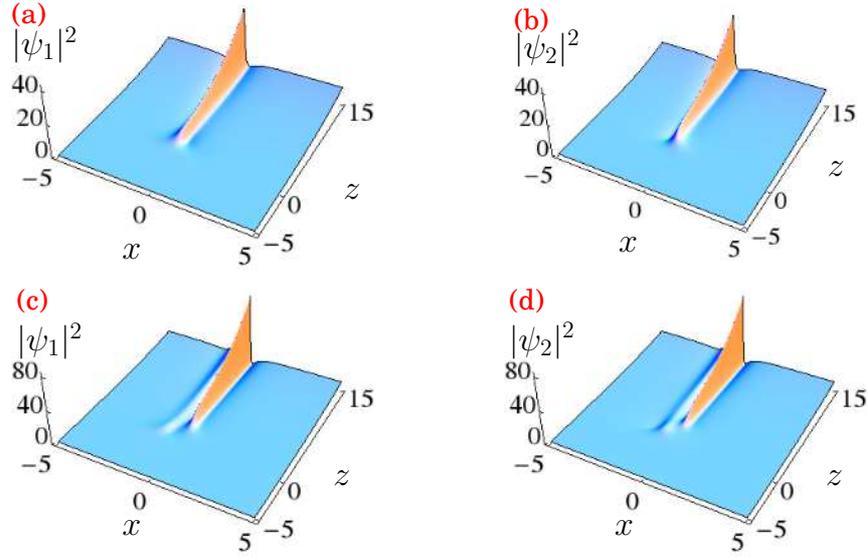}}
\end{center}
\caption{Intensity profiles of the RW for $\beta(z)=\beta_1 \exp{(-\beta_0 z)}$ and $\gamma(z)=\gamma_0/2$.  (a)-(b) Type-I RWs and (c)-(d) Type-II RWs.  The other parameters are same as in Fig. \ref{cdis:fig1} with $\beta_0 = 0.9$, $\beta_1=0.5$}.
\label{cdis:fig2a}
\end{figure*}

\subsection{Case 2}
Next we consider an exponentially distributed diffraction parameter, $\beta(z)=\beta_1 \exp{(-\beta_0 z)}$ with $\gamma(z)=\gamma_0/2$, which has potential applications in the long-distance communication systems \cite{maha}.  Here $\beta_0 > 0$ corresponds to the dispersion decreasing waveguide and $\beta_0<0$ represents the dispersion increasing waveguide.  Substituting this expression into Eq. (\ref{cdis:eq6}), we obtain
\begin{subequations}
\begin{align}
\label{cdis:eq7a}
\psi_j(x,y,z) =  \frac{\beta_0 \rho _0}{a_0 \beta_1+\beta_0 e^{\beta_0 z}} U_j(X,Z)  \eta(x,y,z),\; j=1,2,
\end{align}
\begin{align}
\eta(x,y,z) = & \exp \left(\frac{1}{8} \left(-\frac{4 i b_0 \left(\frac{x}{q_0}+\frac{y}{r_0}\right)}{\frac{a_0 \beta_1 e^{\beta _0 (-z)}}{\beta_0}+1} \right. \right. \notag \\ & \left. \left. +\frac{i \beta_1 b_0^2 \left(q_0^2+r_0^2\right)}{q_0^2 r_0^2 \left(a_0 \beta_1+\beta_0 e^{\beta_0 z}\right)}-\frac{4 i a_0 \left(x^2+y^2\right)}{\frac{a_0 \beta_1 e^{\beta_0 (-z)}}{\beta_0}+1} \right. \right. \notag \\ & \left. \left.  +8 \beta_0 z+4 \gamma_0 z+8 i \phi_0\right)\right) , 
\end{align}
\end{subequations}
where $U_j(X,Z)$'s are given in Eqs. $(\ref{soln})$ and $(\ref{combsoln})$. 

The evolution of Type-I RWs for this diffraction parameter is given in Figs.~\ref{cdis:fig2a}(a)-\ref{cdis:fig2a}(b).  The amplitudes of the RWs in both the components increase monotonically along the propagation direction.  The intensity profiles of Type-II RWs for the same diffraction parameter are presented in Figs.~\ref{cdis:fig2a}(c)-\ref{cdis:fig2a}(d).  Here also the amplitudes of the RWs monotonically increase along the propagation direction.  The width of the RWs does not change.  Thus, choosing an exponentially distributed diffraction parameter we can enhance the amplitude of the optical RWs by keeping the width of them constant.  
\begin{figure*}[!ht]
\begin{center}
\resizebox{0.99\textwidth}{!}{\includegraphics{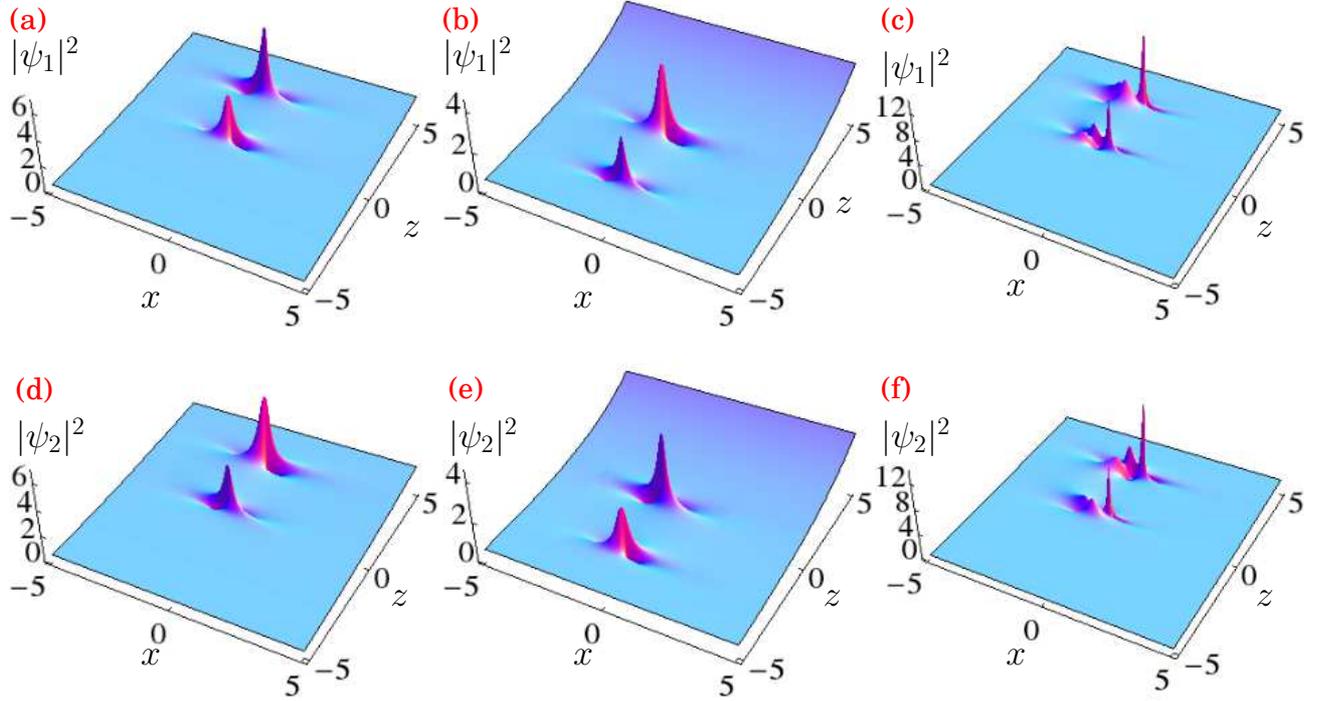}}
\end{center}
\caption{Intensity profiles of composite RWs for $\beta(z)=\beta_1 -\beta_0 z$ and $\gamma(z)=\gamma_0/2$. (a),(b),(d) and (e) Type-I RWs and (c)-(f) Type-II RWs.  The other parameters are same as in Fig. \ref{cdis:fig1} with $\beta_0 = 0.7$, $\beta_1=1$}.
\label{cdis:fig2b}
\end{figure*}
\begin{figure*}[!ht]
\begin{center}
\resizebox{0.75\textwidth}{!}{\includegraphics{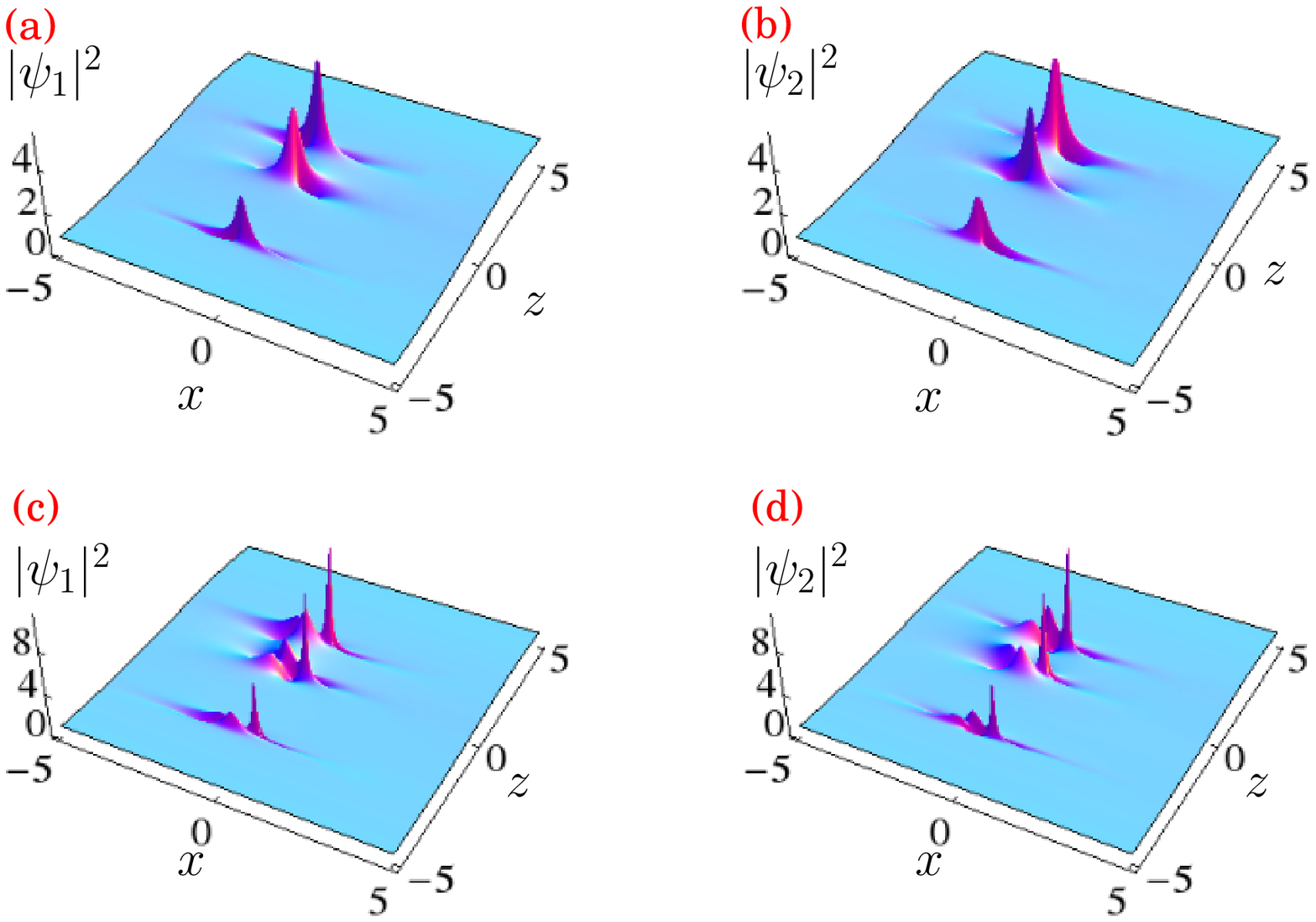}}
\end{center}
\caption{Intensity profiles of composite RWs for $\beta(z)=\beta_1 -\beta_0 z-\beta_2 z^2/2$ and $\gamma(z)=\gamma_0/2$.  (a)-(b) Type-I RWs and (c)-(d) Type-II RWs.  The other parameters are same as in Fig. \ref{cdis:fig1} with $\beta_0 = 0.8$, $\beta_1=2$ and $\beta_2=2.5$}.
\label{cdis:fig2c}
\end{figure*}

\subsection{Case 3}
In the above, we have investigated how the optical RWs get amplified along the propagation direction by considering two different forms of diffraction parameters.  Next we consider a linear profile for $\beta(z)$, that is $\beta(z)=\beta_1-\beta_0 z$ \cite{wang,wang2} with $\gamma(z)=\gamma_0/2$, where $\beta_1$,  $\beta_0$ and $\gamma_0$ are arbitrary parameters.  We construct Type-I and Type-II RW solutions of (\ref{cdis:eq1}) for this profile and analyze how they get modified along the propagation direction.  In the present case, the solution (\ref{cdis:eq6}) reads
\begin{subequations}
\begin{align}
\label{cdis:eq7b}
\psi_j(x,y,z) =  \frac{2 \rho _0}{a_0 s_4 z +2} U_j(X,Z)  \eta(x,y,z), \; j=1,2, 
\end{align}
\scriptsize
\begin{align}
\eta(x,y,z) = & \exp \left(-\frac{i \left(8 b_0 q_0 r_0 \left(q_0 y+r_0 x\right)-b_0^2 z \left(q_0^2+r_0^2\right) s_4\right)}{8 q_0^2 r_0^2 \left(a_0 s_4 z +2\right)} \right. \notag \\ & \left. -\frac{4 i q_0^2 r_0^2 \left(a_0 \left(2 \left(x^2+y^2+\beta _1 z s_3 \right)-\beta _0 z^2 s_3 \right)-2 s_3\right)}{8 q_0^2 r_0^2 \left(a_0 s_4 z +2\right)}\right), 
\end{align}
\scriptsize
\end{subequations}
where $s_3=2 \phi _0-i \gamma _0 z$, $s_4=\beta _0 z-2 \beta _1$, and $U_j(X,Z)$'s are given in Eqs. $(\ref{soln})$ and $(\ref{combsoln})$. 

To begin, we consider Type-I RW solution.  We start our analysis by choosing $\beta_0=0.2$ in the considered profile.  The resultant structure is found to be the same as shown in Figs.~\ref{cdis:fig1}(a)-(b). When we increase the value of $\beta_0$ we obtain composite RWs consisting of two RWs (RW pair).  The outcome is drawn in Fig.~\ref{cdis:fig2b}. When $\beta_0=0.7$, we obtain two separate RWs in each component as shown in Figs.~\ref{cdis:fig2b}(a) and \ref{cdis:fig2b}(d). The RWs occur at two different propagation distances, namely $z_1$ and $z_2$ at $x=0$.  The first RW appears and rapidly disappears around $z_1 \approx 1$ and the second one appears $z_2 \approx 3.5$.  When we increase the value of $\beta_0$ the RWs merge together and when we decrease the value of $\beta_0$ the RWs separate out in the $x-z$ plane which is not shown here. A similar characteristics were also observed in \cite{wang,wang2} in the case of scalar derivative NLS equation. When $\beta_0=-0.7$, the RW pair reemerges in the $x-z$ plane $z \leq 0$ in each component.  The first RW appears and it disappears rapidly at $z_1 \approx -3.5$ whereas the second RW appears at $z_2 \approx 1$ as shown in Figs.~\ref{cdis:fig2b}(b) and \ref{cdis:fig2b}(e). The intensity profiles of Type-II RWs for the same diffraction parameter are displayed in Figs.~\ref{cdis:fig2b}(c) and \ref{cdis:fig2b}(f).  Here also the RWs appear at two different propagation distances, namely $z_1$ and $z_2$ at $x=0$.  When we increase the value of $\beta_0$ the RWs merge together while we decrease the value  of $\beta_0$ the RWs separate out in the $x-z$ plane.  When $\beta_0=-0.7$, the RW pair reemerges in the plane $z \leq 0$ in each component.  We also observe from these figures that the amplitude of the RWs increases in $z_1 < z_2$ manner in the $x-z$ plane.  As far as the linear profile is concerned our results reveal that the relative position of one of the two constituents of RWs can be altered.
\subsection{Case 4}
Finally, we choose the diffraction parameter in the form $\beta(z)=\beta_1 -\beta_0 z-\beta_2 z^2/2$ with $\gamma(z)=\gamma_0/2$, where $\beta_1$,  $\beta_0$, $\beta_2$ and $\gamma_0$ are arbitrary parameters \cite{wang}.  Substituting this diffraction parameter into Eq. (\ref{cdis:eq6}), we obtain
\begin{subequations}
\label{cdis:eq7c}
\begin{align}
\psi_j(x,y,z) = & \, \frac{6 \rho _0}{\left(a_0 s_5 z +6\right)}  \times U_j(X,Z) \eta(x,y,z), \; j=1,2, 
\end{align}
\scriptsize
\begin{align}
\eta(x,y,z) = & \exp \left(-\frac{i \left(24 b_0 q_0 r_0 \left(q_0 y+r_0 x\right)-b_0^2 z \left(q_0^2+r_0^2\right) s_5 \right)}{8 q_0^2 r_0^2 \left(a_0 z s_5+6\right)} \right. \notag \\  & \left. -\frac{i \left(a_0 \left(6 (x^2+ y^2)+i \beta _2 \gamma _0 z^4-s_6 \right)-6s_3\right)}{2 \left(a_0 z s_5+6\right)}\right), 
\end{align}
\normalsize
\end{subequations}
where $s_5=-6 \beta _1+\beta _2 z^2+3 \beta _0 z$, $s_6=2 \beta _2 z^3 \phi _0+3 \beta _0 z^2 s_3 -6 \beta _1 z s_3$ and $U_j(X,Z)$'s are given in Eqs. $(\ref{soln})$ and $(\ref{combsoln})$. 

This choice provides the composite RWs consist of three RWs in both types.  The triplet RWs appear along the propagation direction for $\beta_0 = 0.8$, $\beta_1=2$ and $\beta_2=2.5$ as shown in Fig. \ref{cdis:fig2c}.  The intensity profiles of both Type-I and Type-II RWs (composite RWs) of  each component are shown in Figs. \ref{cdis:fig2c}(a)-(b) and Figs. \ref{cdis:fig2c}(c)-(d), respectively.  We observe that three separate RWs occur at three different propagation distances, namely $z_1$, $z_2$ and $z_3$ and their amplitudes increase in the order $z_1 < z_2 < z_3$ in the $x-z$ plane. For this type of diffraction parameter, we conclude that the location of the composite RWs can be manipulated.  Very recently an attempt has been made to transform the two-dimensional two coupled vcNLS equations into two coupled constant coefficient NLS equations \cite{wang3}. In this work, the authors have identified only the vector Peregrine soliton solution and bright-dark-soliton-RW solutions.

\section{Characteristics of dark-dark RWs}
In this section, we pay our attention on identifying another localized structure, namely dark-dark RWs in the two coupled vcNLS equations.  To construct this localized solution, we consider the (2+1)-dimensional defocusing coupled vcNLS equations \cite{book1,book2}, that is  
\begin{align}
i\psi_{1z}+\frac{\beta(z)}{2}\left(\psi_{1xx}+\psi_{1yy}\right)-R(z)\sum_{k=1}^{2} \vert \psi_k \vert^2 \psi_1 = i \gamma(z)\psi_1, \nonumber \\
i\psi_{2z}+\frac{\beta(z)}{2}\left(\psi_{2xx}+\psi_{2yy}\right)-R(z)\sum_{k=1}^{2} \vert \psi_k \vert^2 \psi_2 = i \gamma(z)\psi_2,
\label{dcdis:eq1}
\end{align}
where $\psi_1(x,y,z)$ and $\psi_2(x,y,z)$ denote the complex envelope of the electrical fields in the moving frame, $z$ is the coordinate along the propagation direction and $x$ and $y$ are the transverse coordinates.  The function $\beta(z)$ represents the diffraction coefficient, $R(z)$ is the nonlinearity parameter and $\gamma(z)$ is the gain or loss parameter.  Equation (\ref{dcdis:eq1}) can be transformed to the defocusing two coupled NLS equation (\ref{cnls1}) through the similarity transformation (\ref{cdis:eq2}). The unknown functions $\rho(z)$, $\phi(x,y,z)$ and $X(x,y,z)$ can be determined by following the same procedure as given in Sec. 2.  Repeating the steps we obtain the same expressions as given in (\ref{cdis:eq3a})-(\ref{cdis:eq3d}) except $Z(z)$ which reads now $Z(z)= -\frac{(q_0^2+r_0^2)M(z)}{2f_0^2\epsilon(1-a_0 M(z))}$. The existence of dark RW solutions in (\ref{dcdis:eq1}) can be guaranteed by the fulfillment of the same constraint (\ref{cdis:eq4}).  The only  difference which we make here is that the functions $U_j(X,Z)'$s should satisfy the coupled NLS equations of the form
\begin{eqnarray}
\label{cnls1}
i \frac{\partial U_1}{\partial Z}-\frac{\partial ^2 U_1}{\partial X^2}+ 2 (|U_1|^2+|U_2|^2) U_1=0, \nonumber \\
i \frac{\partial U_2}{\partial Z}-\frac{\partial ^2 U_2}{\partial X^2}+ 2 (|U_1|^2+|U_2|^2) U_2=0,
\end{eqnarray}
\begin{figure*}[!ht]
\begin{center}
\resizebox{0.75\textwidth}{!}{\includegraphics{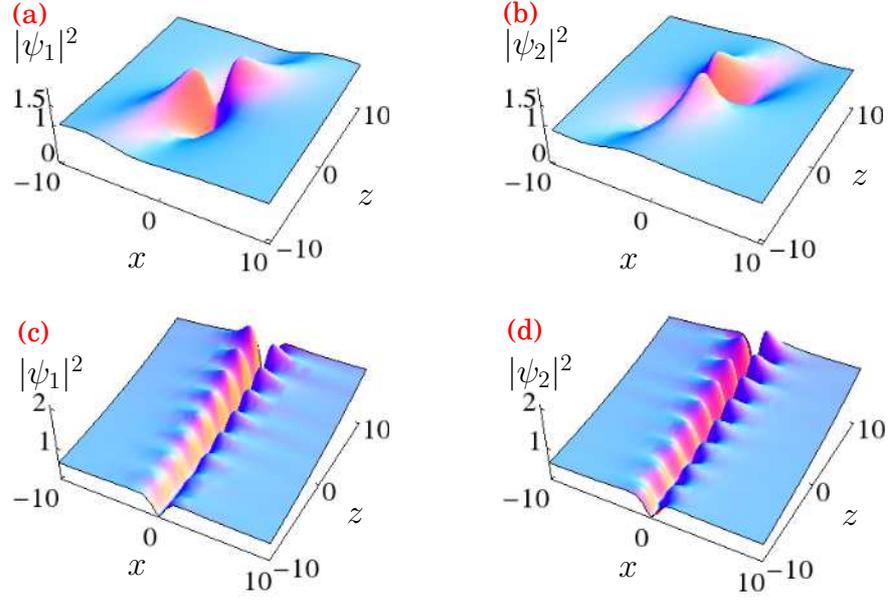}}
\end{center}
\caption{Intensity profiles of the dark-dark RWs for $\beta(z)=\beta_0 \cos{(\sigma z)}\exp{(\gamma_0 z)}$ and a gain/loss parameter $\gamma(z)=\gamma_0/2$.  When the parameter $\gamma_0 =0$ and $\sigma =0$ for (a)-(b) and $\gamma_0=0.1$ and $\sigma=2.5$ for (c)-(d).  The other parameters are same as in Fig. \ref{cdis:eq1} with $h=1$, $s=0.5$.}
\label{dcdis:fig1}
\end{figure*}
\begin{figure*}[!ht]
\begin{center}
\resizebox{0.95\textwidth}{!}{\includegraphics{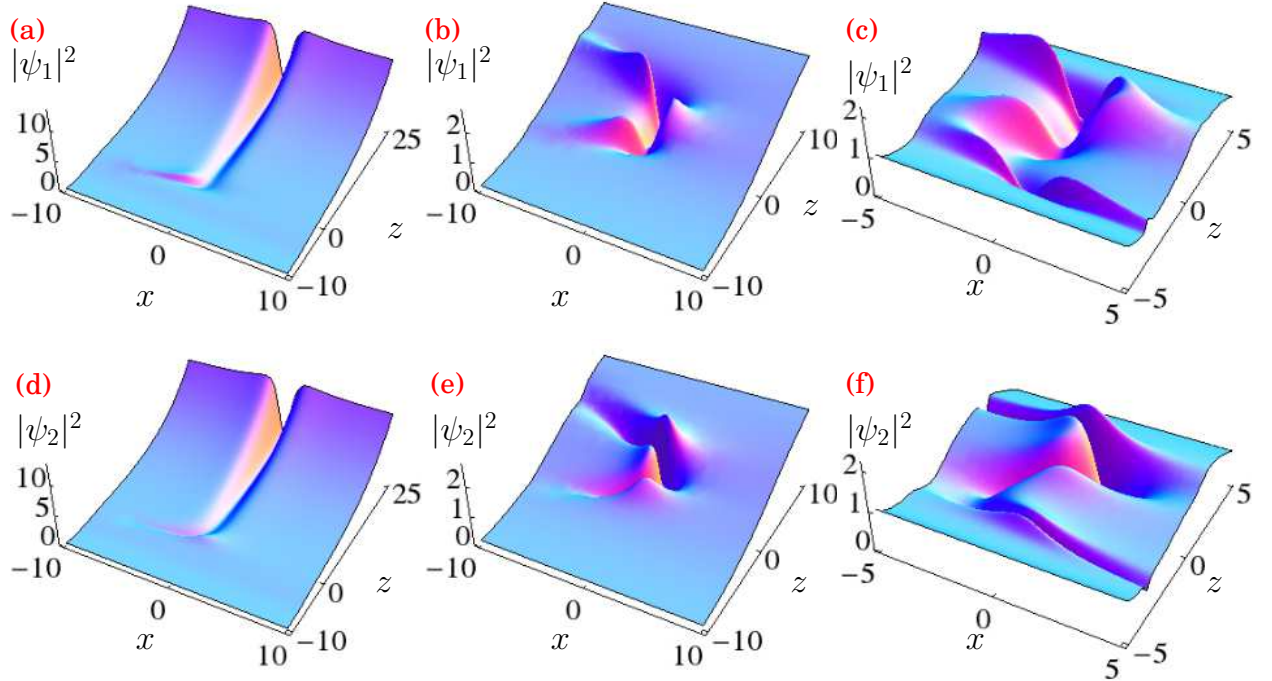}}
\end{center}
\caption{Intensity profiles of the dark-dark RWs for (a), (d) $\beta(z)=\beta_1 \exp{(-\beta_0 z)}$, (b), (e) $\beta(z)=\beta_1-\beta_0 z$ and (c), (f) $\beta(z)=\beta_1 -\beta_0 z-\beta_2 z^2/2$.  The other parameters are same as in Fig. \ref{cdis:eq1} with $h=1$, $s=0.5$.}
\label{dcdis:fig2}
\end{figure*}
which is different from (\ref{cnls}). Here the nonlinearity and dispersion are in opposite signs.  The existence of RWs in the normal dispersion regime of two coupled NLS equation (\ref{cnls1}) has been investigated in \cite{lomb,lomb1}. Vector RW (black or dark RW) solutions of (\ref{cnls1}) have been constructed through Hirota bilinear transform method \cite{jhli}.  It has been found that the resultant RW is localized algebraically in both space and time similar to the Peregrine soliton of a scalar NLS equation \cite{pere}.  The difference occur at the center of the RW where the amplitude is less than the amplitude of the background. This localized structure is called dark RW. 

Equation (\ref{cnls1}) admits the following form of dark-dark RW solutions \cite{jhli} 
\begin{eqnarray}
\label{darksoln}
U_1(X,Z)&=& h \exp{(4 i h^2 Z)} \notag \\ && \, \left(1+\frac{4 g^2(-1+i(-s X-s^2 Z + g^2 Z))}{(g^2+s^2)(g^2(X+s Z)^2+g^4Z^2+1)}\right), \notag \\
U_2(X,Z)&=& h \exp{(i s X+i s^2 Z+4 i h^2 Z)} \\ && \, \left(1+\frac{4 g^2(-1+i(-sX-s^2 Z+g^2 Z))}{(g^2+s^2)(g^2(X+s Z)^2+g^4Z^2+1)}\right),\nonumber
\end{eqnarray}
where $g=\pm \sqrt{-4h^2-s^2+2\sqrt{(2 h^2)^2+4s^2h^2}}$, $h$ denotes the amplitude of the continuous wave background and $s$ is an arbitrary parameter and the criterion for the existence of RW is $s^2<8h^2$.  

As long as the condition (\ref{cdis:eq4}) is satisfied, the exact solution of Eq. (\ref{dcdis:eq1}) can be given in the form 
\begin{eqnarray}
\label{dcdis:dark}
\psi_j(x,y,z)&=&\frac{\rho_0\exp(G(z))}{1-a_0M(z)}U_j(X,Z) \exp \left[i\left(\phi_0 \right. \right. \notag \\ && \left. \left. -\frac{a_0(x^2+y^2)}{2(1-a_0M(z))}-\frac{b_0(x/q_0+y/r_0)}{2(1-a_0M(z))} \right. \right. \notag \\ && \left. \left. -\frac{(q_0^2+r_0^2)b_0^2M(z)}{8q_0^2r_0^2(1-a_0M(z))}\right)\right],  \; j=1,2,
\end{eqnarray} 
where $U_j(X,Z)'s$, is the solution of defocusing coupled NLS equations which is given in Eq. (\ref{darksoln}). 
In the following, we discuss in detail how the dark-dark RW structures get deformed for different forms of diffraction parameter $\beta(z)$.

As we done earlier, to begin, we consider an exponentially growing periodic diffraction parameter, say $\beta(z)=\beta_0 \cos{(\sigma z)}\exp{(\gamma_0 z)}$ with $\gamma(z)=\gamma_0/2$.  Substituting these expressions into (\ref{dcdis:dark}), we can obtain the dark-dark RW solutions of (\ref{dcdis:eq1}).  The intensity profiles of the dark-dark RWs of (\ref{dcdis:eq1}) are depicted in Fig. \ref{dcdis:fig1}.   The intensity profiles of the constant dispersion parameter $\beta(z)=\beta_0$ ($\sigma, \gamma_0=0$) are shown in Fig.~\ref{dcdis:fig1}(a)-\ref{dcdis:fig1}(b), affirms the fundamental RW features.  As we visualize the amplitude of the dark RW of first and second component turns out to be the same even though they are localized in different orientations.  For the parametric choice $\gamma_0=0.1$ and $\sigma=2.5$, the dark RW in each component propagates periodically and their intensities increase gradually along the propagation direction as shown in Figs.~\ref{dcdis:fig1}(c)-\ref{dcdis:fig1}(d).  From the outcome, we conclude that the dark RW is not being distorted for the constant diffraction parameter whereas in the case of the exponentially growing periodic diffraction parameter the dark RWs of both the components exhibit breathing nature along the propagation direction. 

Next we consider an exponentially distributed diffraction parameter $\beta(z)=\beta_1 \exp{(-\beta_0 z)}$, where $\beta_1$ and $\beta_0$ are arbitrary parameters \cite{maha}.  The intensity profiles of the modified dark-dark RWs for this diffraction parameter are shown in Figs. \ref{dcdis:fig2}(a) and \ref{dcdis:fig2}(d).  The amplitudes of the dark RWs are equal but the dips of the modified dark RWs become more darkened in both the components. We also observe that the background of the RW becomes steeper due to the presence of gain parameter $\gamma_0$ which is chosen as $0.15$. When we increase the value of $\gamma_0$, the background gets more steeper and when we decrease the value of $\gamma_0$ the background becomes flatter in both the components.

The characteristics of dark-dark RWs (\ref{dcdis:dark}) which come out from the linear diffraction parameter, that is $\beta(z)=\beta_1-\beta_0 z$ and $\gamma(z)=\gamma_0/2$, where $\beta_0$, $\beta_1$ and $\gamma$ are arbitrary parameters differ from the above two cases. With $\beta_0=0.7$, we come across an interaction of two separate dark RWs in each component as shown in Figs. \ref{dcdis:fig2}(b) and \ref{dcdis:fig2}(e). The amplitudes of the dark RW pair of first and second component are found to be equal but they are localized in different positions.  The two dark RWs occur at two different propagation distances, namely $z_1$ and $z_2$, in the $x-z$ plane. When we increase the value of $\beta_0$ we observe that the RW pair separates out from each other and when we decrease the value $\beta_0$ the RW pair merges together in the plane wave background. Here also when we increase the value of $\gamma_0$, the background gets more steeper and when decrease the value of $\gamma_0$ the background becomes flatter in both the components.

The intensity profiles of three composite dark RWs which arise in individual components for $\beta(z)=\beta_1 -\beta_0 z-\beta_2 z^2/2$ are demonstrated in Figs. \ref{dcdis:fig2}(c) and \ref{dcdis:fig2}(f). The amplitudes of the dark RWs of first and second component are equal but they are localized in different locations.  The three dark RWs occur at three different locations, namely $z_1$, $z_2$ and $z_3$, in the $x-z$ plane in each component. Here also we come across a bending background due to presence of gain parameter. When we increase the value of $\gamma_0$ from $0.15$, the curved background gets more steeper whereas when we decrease the value of $\gamma_0$ the background becomes flatter in both the components.

\section{Conclusion}
In this work, we have constructed vector RW solutions for the two-dimensional two coupled NLS equations with distributed coefficients such as diffraction, nonlinearity and gain/loss parameter.  We have transformed the vcNLS equations into the Manakov equation through similarity transformation with a constraint. The Manakov equation admits a family of RW solutions.  For our studies we have considered two types of RW solutions.  By substituting these two RW solutions back in the similarity transformation we have obtained two different types of RW solutions for the considered equation. We then investigated the characteristics of these two types of RW solutions by considering four different forms of diffraction parameters.  The outcome was reported in detail.  We have also constructed the dark-dark RW solutions for the coupled two-dimensional inhomogeneous NLS equations and investigated the characteristics of dark RWs for the same forms of diffraction parameters.  We have observed some novel characteristics among the vector RWs. Our results will be useful for the experimental research on fiber lasers and super continuum generation.

\begin{acknowledgement}
KM thanks the University Grants Commission (UGC-RFSMS), Government of India, for providing a research fellowship. The work of MS forms part of a research project sponsored by NBHM, Government of India. 
\end{acknowledgement}
All the authors have contributed equally to the research and to the writing up of the paper.

\end{document}